\documentclass[twocolumn,showpacs,preprintnumbers,amsmath,amssymb]{revtex4}
	\usepackage{mathrsfs}
	\usepackage{graphicx}
	\usepackage{float}

\begin{document}
	\title{\textbf{Complementarity of Dark Matter Searches at Resonance}}
\author{Malcolm Fairbairn}
 \email{malcolm.fairbairn@kcl.ac.uk}
\affiliation{%
Department of Physics, King's College London, Strand, London WC2R 2LS, UK
}%
\author{John Heal}
\email{john.heal@kcl.ac.uk}
\affiliation{%
Department of Physics, King's College London, Strand, London WC2R 2LS, UK
}%

	\date{June 12, 2014}
	\newcommand{\ma}{m_{A^\prime}}

\begin{abstract}
We consider models of dark matter where the couplings between the standard model and the dark sector fall at resonance due to kinematics and direct detection experiments become insensitive.  To be specific, we consider a simple model of 100 GeV - TeV scale dark matter coupled to the standard model via a vector boson. 
We explore whether it will be possible to exclude such regions of the parameter space using future observations of dijet rates at the LHC and CTA and AMS observations of the Galactic Centre.
\end{abstract}

\maketitle

\section{Introduction}
One of the most compelling classes of dark matter are the thermal relics (including weakly interacting massive particles) where self-annihilation cross sections set by new physics at the electroweak scale give rise to a density comparable to that observed in the Universe today.  There are many experimental searches in progress looking for this kind of dark matter. Collider searches (such as the LHC and its predecessor the Tevatron) hope to produce heavy dark matter particles directly in a laboratory setting and to test the interactions of heavier short lived particles to look for signals of new mediators to the dark sector. There are also direct detection experiments, such as LUX \cite{Akerib:2013tjd}, looking for the interaction of astrophysical dark matter with baryons in detectors.  Finally there are also indirect searches for dark matter such as Fermi \cite{Ackermann:2013yva} and CTA \cite{Doro:2012xx} which search for the standard model products associated with the self-annihilation of dark matter in space.

Each of these different methods for searching for dark matter adds complementary information which is very important in finding out if a given dark matter model is excluded by experimentation or not. One classic example is that of a fermionic dark matter particle which interacts with the rest of the standard model via an additional boson which has not yet been discovered, for example a new vector boson (see next section). The relic abundance of the dark matter particle is set by the rate at which it annihilates with itself in the early Universe, which depends upon both gauge couplings and kinematics. If the masses of the new particles just happen to be such that the dark matter has close to half the mass of the gauge boson then the annihilation will be resonant and the couplings can be small to achieve the correct relic abundance. 

However if the couplings are small then the scattering of the dark matter particle off nuclei in detectors will also be suppressed.  As the dark matter-nucleon cross section goes to lower cross sections it will eventually be so small that the number of nuclear recoils from neutrinos in a given experiment will create a large irreducible background \cite{Billard:2013qya}, so in those regions it is paramount that either LHC constraints or indirect detection constraints can step in and give discriminating power.

The resonance will however increase the probability of the dark matter particle annihilating with itself, so we increase the sensitivity of gamma ray experiments such as CTA and Fermi for detecting a signal from this candidate.

In this paper we explore the complementarity of various experimental results on a simplified dark matter model, specifically to explore our ability to rule out or detect the model at resonance. First in Sec. \ref{setup} the model is introduced, then in Sec. \ref{exp} the various experimental constraints on dark matter are considered separately in the case where we keep the mediator mass fixed in order to understand qualitatively the way different constraints apply. Afterwards in Sec. \ref{combine} the different constraints are combined both for the case where we have a single mediator mass and when we allow it to vary.  Finally we will discuss the results.

\section{The Theoretical Setup\label{setup}}
\subsection{Lagrangian}
We consider an extension to the standard model where a Dirac fermion dark matter candidate $\chi$ couples to the Standard Model through a new massive vector boson $A^\prime$ (with a field strength $F^\prime$).  This additional vector boson couples to each of the quarks in a flavour blind way - the modification to the Lagrangian is given in equation~\ref{eq:lagrangian} (this Lagrangian has been previously studied elsewhere \cite{Graesser:2011vj,Shoemaker:2011vi,Frandsen:2012rk,Buchmueller:2013dya,Arcadi:2013qia,Buchmueller:2014yoa,Malik:2014ggr}): 
\begin{align}
 \Delta\mathscr{L}=&-\frac{1}{4}F^{\prime}_{\mu\nu}F^{\prime\mu\nu} +\frac{1}{2}\ma^2 A^{\prime}_\mu A^{\prime\mu}\nonumber\\ 
&+ \bar{\chi}(\gamma^{\mu}\partial_\mu - m_{\chi})\chi+A^{\prime}_{\mu}\bar{\chi}\gamma^{\mu}(g_{\chi V} - g_{\chi A}\gamma^5 )\chi\nonumber\\
&+  A^{\prime}_{\mu}\bar{q}\gamma^{\mu}(g_{q V} - g_{q A}\gamma^5 )q\label{eq:lagrangian}
\end{align}

Without specifying the new Beyond the Standard Model (BSM)  physics theory the direct mass for $A^\prime$ is not gauge invariant but we assume the mass arises due to new physics at higher energies. The mediator mass is fixed for the first part of this analysis to $\ma=3$ TeV in order to study in detail the behaviour of couplings around the resonance more clearly.  We choose 3 TeV as it will be an energy range where dijet information is likely to improve significantly. We chose to focus on the pure vector case ($g_{\chi A}, g_{qA}=0$) and the couplings are allowed to vary between $0<g_{\chi V},g_{qV}<3$  to maintain perturbativity while we consider dark matter masses in the range $0<m_\chi<2.5$ TeV. 
The decay width of $A^\prime$ is given by
\begin{align}
	\Gamma_{A^\prime}=&\frac{g_{\chi V}m_{A^\prime}}{12\pi}\sqrt{1-\frac{4m_\chi^2}{\ma^2}}\left(1+\frac{2m_\chi^2}{\ma^2} \right)\label{eq:decaywidth} \\
										&+\sum_q\frac{N_cg_{qV}\ma}{12\pi}\sqrt{1-\frac{4m_q^2}{\ma^2}}\left(1+\frac{2m_q^2}{\ma^2}\right)\nonumber
\end{align}
where we respect $\Gamma_{A^\prime}<0.15\ma$ in order to allow all the points to be studied in the context of CMS's narrow dijet searches discussed in Sec.~\ref{sec:dijet}. This constraint places a limit on $g_{qV}$ while leaving $g_{\chi V}$ unaffected as the terms in Eq.~\ref{eq:decaywidth} which are related to $g_{qV}$ come with an extra factor of the number of colours and flavours of quarks that the mediator can decay to, so small increases to $g_{qV}$ increase $\Gamma_{A^\prime}$ more rapidly than an equivalent change in $g_{\chi V}$. The majority of the couplings which give $\Gamma_{A^\prime}>0.15\ma$ have a product of the coupling which would lead to a direct detection cross section (discussed in Sec.~\ref{sec:direct}) which is already excluded by experimental searches.

\subsection{Relic density}
The density of dark matter observed today assuming the normal $\Lambda$CDM cosmology is set by observations from the Planck experiment \cite{Ade:2013zuv} which when combined with WMAP \cite{WMAP9} observations give (where $h$ is the dimensionless Hubble's constant)
\begin{equation}
	\Omega_{DM}h^2=0.1198\pm0.0026
\end{equation}

While the relic density could be provided by a combination of particles, here $\chi$ is treated as the only stable BSM particle so it must provide the full relic density within 2$\sigma$ of the Planck value.  We assume that $\chi$ is in thermal equilibrium with the rest of the plasma at early times and that the relic density is given by the comoving density of particles after freeze-out of the equilibrium.

We use \textsc{micrOMEGAs} \cite{Belanger:2013oya} to calculate the relic density for the model considered here using the full integral formulation to avoid potential pitfalls in velocity expansion methods  which could become more significant in resonance regions \cite{Gondolo:1990dk}.
The cross section involved in the annihilation which determines abundance at the point of thermal freeze-out can be written

\begin{eqnarray}
	\sigma(s)&=&\frac{4g_{qV}^2g_{\chi V}^2 s}{3\pi (s-\ma^2)^2+\ma^2\Gamma_{A^\prime}^2}\sum_q\frac{\beta_q}{\beta_\chi}(1\nonumber\\
						&&+\frac{2m_\chi^2}{s}+\frac{2m_q^2}{s}+\frac{16m_\chi^2m_q^2}{s^2})
	\label{eq:relicM}
\end{eqnarray}

where $s$ is the centre of mass energy and the parameters $\beta_q$ and $\beta_\chi$ are given by

\begin{equation}
  \beta_i=\sqrt{1-\frac{4m_i^2}{s}}
\end{equation}
Most of the energy of the annihilating dark matter particles is in their rest mass at freeze-out, so $s\approx 4m_\chi^2$. Because of this, around $2m_\chi \approx m_{A'}$ the product of the couplings goes towards zero. This is due to the cross section rapidly increasing as the annihilation approaches the resonant production of the mediating $A^\prime$, so to keep with the range for the relic density $g_{\chi V}^2 g_{qV}^2$ has to become correspondingly smaller. 
\section{Separate Experimental Constraints}
\label{exp}
\subsection{Direct detection}
\label{sec:direct}
Recently liquid xenon detectors have given the best constraints on dark matter nucleon interactions using a combination of scintillation and ionization to help further discriminate between background and signal events by providing greater identification of nuclear recoil events.

\begin{figure}[t]
	\centering
	\includegraphics[width=0.5\textwidth]{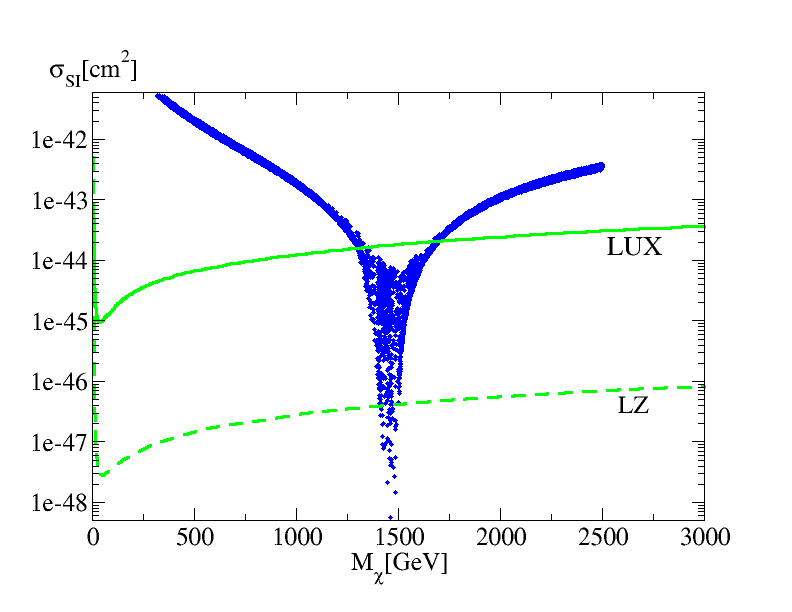}
	\caption{\textit{Current LUX and estimated LZ constrains on the spin-independent cross section as a function of mass (solid and dashed green lines respectively) with the points which have a narrow $A^\prime$ and satisfy the relic density criteria.\label{fig:direct}}}
\end{figure}

Here the pure vectorial case ($g_{\chi A}=g_{qA}=0$) is investigated such that the spin independent direct detection experiments provide the strictest limits of which LUX \cite{Akerib:2013tjd} is the most recent and has the greatest exclusion for the dark matter mass range of interest (the exclusion is shown in Fig.~\ref{fig:direct}). Alternatively the axial coupling could be investigated to find the impact of spin dependent searches on such models \cite{Archambault:2012pm,Felizardo:2011uw,Behnke:2012ys} as well as monojet searches \cite{Buchmueller:2013dya}.  For the low energies of the dark matter nucleon interaction with respect to the chosen mediator mass, resonance effects can be ignored. Even taking into account the projected future sensitivity of direct detection experiments (such as LUX-ZEPPELIN \cite{Malling:2011va}) these effects can produce thermal dark matter which would be unseen in direct detection.

\subsection{Narrow dijet resonance}

This model naturally leads to changes in the rate of dijet production at the LHC since there are $q\bar{q}\rightarrow A^{\prime}\rightarrow q\bar{q}$ processes, which would be produced resonantly.  The use of dijets to constrain dark matter and the complementarity of such approaches with other search methods has been studied for a variety of models \cite{Frandsen:2012rk,Park:2009cs,Tsai:2013bt,Dutta:2014kia,An:2012va,Alves:2013tqa}. Though a signal alone would not be a clear sign of dark matter, it could be used as a cross-check for models which avoid the LUX constraint. Current CMS dijet limits on the product of the production cross section, detector acceptance and branching ratio for the decay into quark pairs for new physics \cite{CMS:dijet13} (ATLAS has also searched for signals in this channel \cite{ATLAS:2012qjz}) do constrain some of the parameter space. 

\begin{figure}[t]
	\center
	\includegraphics[width=0.5\textwidth]{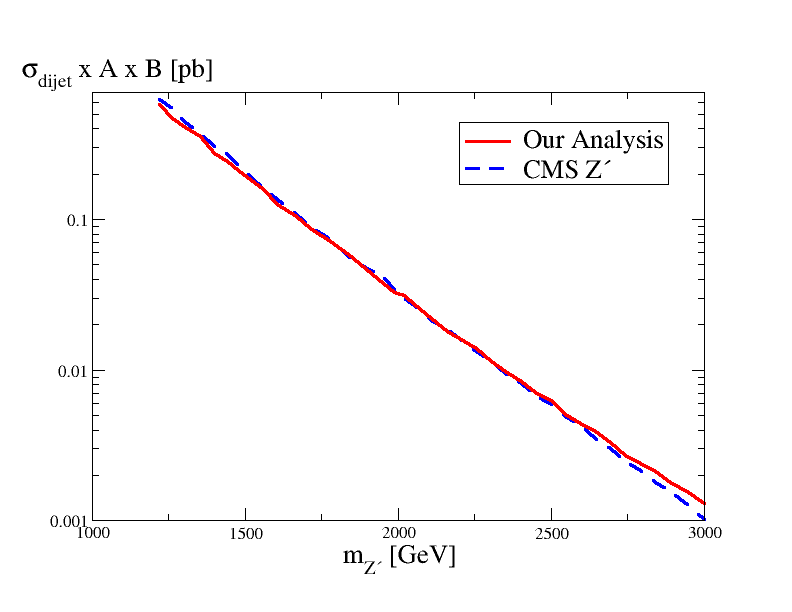}
	\caption{\textit{Predicted dijet cross section of a $Z^{\prime}$ factoring in the $Z^{\prime}$ branching ratio to quarks and the detector acceptance after applying the CMS kinematic cuts calculated using the same analysis method we used to study the simplified dark matter model (red solid line). The $Z^{\prime}$ cross section stated in the CMS analysis (blue dashed line) is shown for comparison.\label{fig:Zp_compare}}}
\end{figure}

The cross sections and the branching ratio for this model were calculated using \textsc{Madgraph5} \cite{MG5} and then event generation and showering were completed with \textsc{Pythia} \cite{pythia8,pythia6} to find the acceptance for each point. The outgoing jets were formed using the anti-$k_T$ algorithm  in \textsc{FastJet} \cite{FastJet} using the trigger criteria from the CMS analysis as the \textsc{FastJet} bounds ($p_T > 30$ GeV and $|\eta|<2.5$). Events with fewer than two jets meeting this criteria were ignored and the lower limit on the invariant dijet mass (defined in the CMS paper, $m_{jj}>890$ GeV) and the upper bound of the pseudorapidity separation of the jets ($|\eta_{jj}|<1.3$) were applied to the two highest $p_T$ jets. To check the validity of this method we ran the same analysis on one of the benchmark models used in the CMS paper (a $Z^\prime$ model \cite{supercolphys}) and find it is in good agreement with the new physics cross section limit shown in their work which can be seen in Fig.~\ref{fig:Zp_compare}.

\begin{figure}[t]
  \center
  \includegraphics[width=0.5\textwidth]{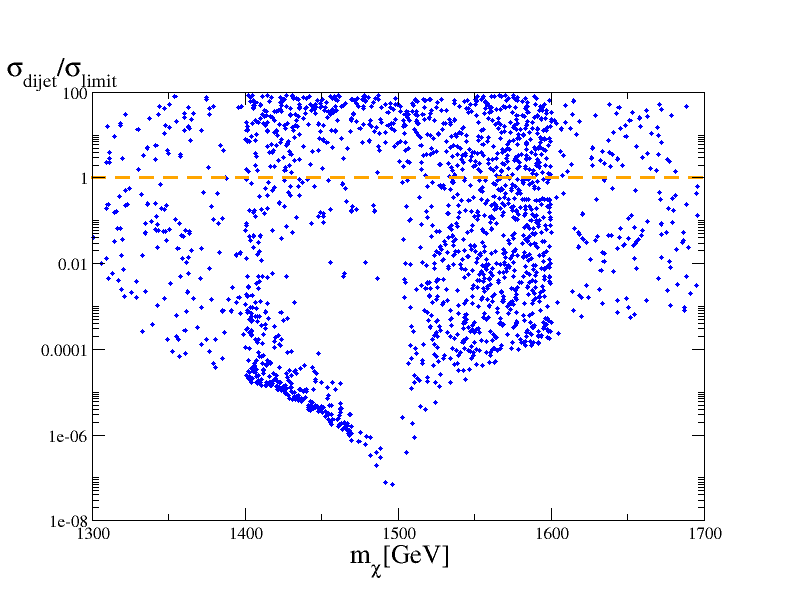}
  \caption{\textit{Dijet limits and expected exclusions from a 14 TeV LHC with 300 $\rm{fb^{-1}}$ as the ratio of the dijet resonance cross section and the CMS dijet search limit as a function of $m_\chi$. The vertical structure in the density of points at $m_\chi=1400$ GeV and $1600$ GeV is a numerical artefact of the search strategy\label{fig:dijetmass}}}
  \end{figure}

As mentioned earlier, to be able to use the dijet limit we have restricted the decay width of our mediator to $\Gamma_{A^\prime} < 0.15 \ma$. While the t-channel exchanges are allowed their contributions are minimal compared to the s-channel on the resonance. To extend the effectiveness of this search channel, we estimate the limits after the 14 TeV run of the LHC producing 300 $\rm{fb^{-1}}$ of data assuming no changes to the analysis and a similar signal to background ratio. This means the major increase in sensitivity would be from the increase in luminosity, so the experimental limits are scaled as a the ratio of the square roots of the luminosity. Figure~\ref{fig:dijetmass} shows the effectiveness of the dijet searches in excluding parameters which would not be seen by LUX.  This plot also shows an asymmetry around the resonance - a mediator produced on shell can decay into a dark matter pair as $2m_\chi < \ma$, which gives another decay channel for the mediator and thus reduces the branching ratio to dijets.  The dijet limit is therefore stronger on the right of the resonance.

	~
	\begin{figure}[t]
	\center
	\includegraphics[width=0.5\textwidth]{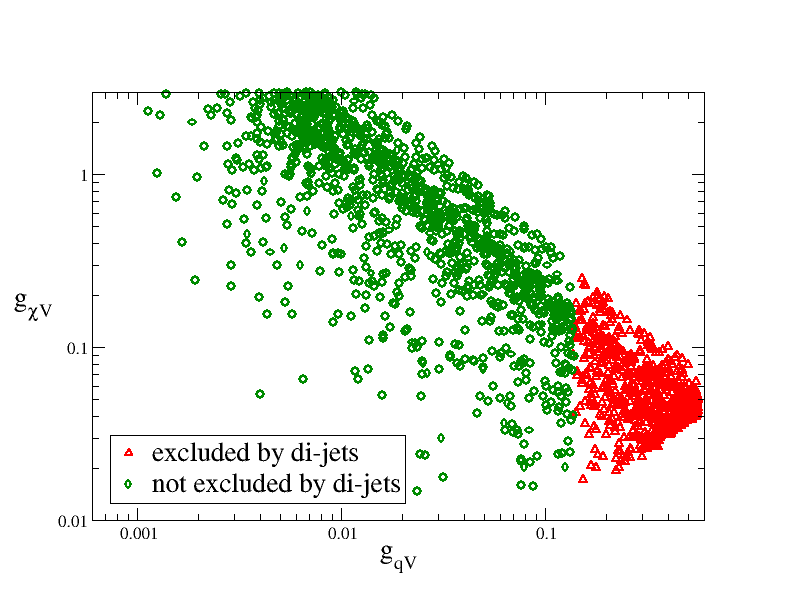}
	\caption{\textit{Dijet limits and expected exclusions from a 14 TeV LHC with 300 $\rm{fb^{-1}}$ shown as the couplings of the points which pass LUX, red triangles excluded by 14 TeV dijets. \label{fig:dijetcoupling}}}
\end{figure}
\label{sec:dijet}

Looking at the effect this has on the allowed couplings (Fig.~\ref{fig:dijetcoupling}) it is clear that the dijet searches can put a strict constraint on $g_{qV}$ but have no direct effect on $g_{\chi V}$. This arises from the the major diagram in this model being proportional to $g_{qV}^4$ unlike both the relic density and direct detection diagrams which scale as $g_{qV}^2g_{\chi V}^2$. Points with high $g_{\chi V}$ are favoured as  the relic density fixes a value for the product of the couplings for a given set of masses, and LUX only constrains this product.

While monojet and $\displaystyle{\not}E_T$ searches are the favoured dark matter detection channel for collider searches (and have been studied in depth for this simplified model elsewhere \cite{Shoemaker:2011vi,Buchmueller:2013dya,Buchmueller:2014yoa}) they typically set a constraint on the product $g_{qV}^2g_{\chi V}^2$. For the parameters which satisfy the dijet width constraints and evade detection by LUX the values of $g_{qV}^2g_{\chi V}^2$ would produce a monojet signature at least 2 orders of magnitude below the current discovery bound from the most recent CMS results \cite{CMS:monjet13}.

\subsection{Indirect detection}

Finally we look at the implications of the indirect limits that can be obtained from the self-annihilation of dark matter in regions of high density. The detectors are set up to observe the distribution of high energy photons, positrons and antiprotons from these proposed dense regions.

For gamma rays, the future Cherenkov Telescope Array (CTA) will be a promising experiment \cite{Acharya:2013sxa,Ripken:2012db,Consortium:2010bc} and many studies into the use of the CTA in dark matter detection have already been carried out \cite{Conrad:2012ju,Bergstrom:2013uxa,Funk:2013gxa,Oakes:2013aya,Bertone:2011pq,Doro:2012xx,Roszkowski:2014wqa,Pierre:2014tra,Wood:2013taa}. 

We also look at constraints from the AMS experiment on the production of antiprotons which can offer a useful constraint for a specific kinematic region \cite{AMS,cirrelliantiprot}.  We use \textsc{micrOMEGAs} to first calculate the self-annihilation cross section for $\chi\bar{\chi}$ to $\Sigma$q$\rm\bar{q}$ in the Galactic centre and compared it to the proposed CTA \cite{Wood:2013taa} and AMS limits \cite{cirrelliantiprot}. 

\begin{figure}[t]
	\centering
	\includegraphics[width=0.5\textwidth]{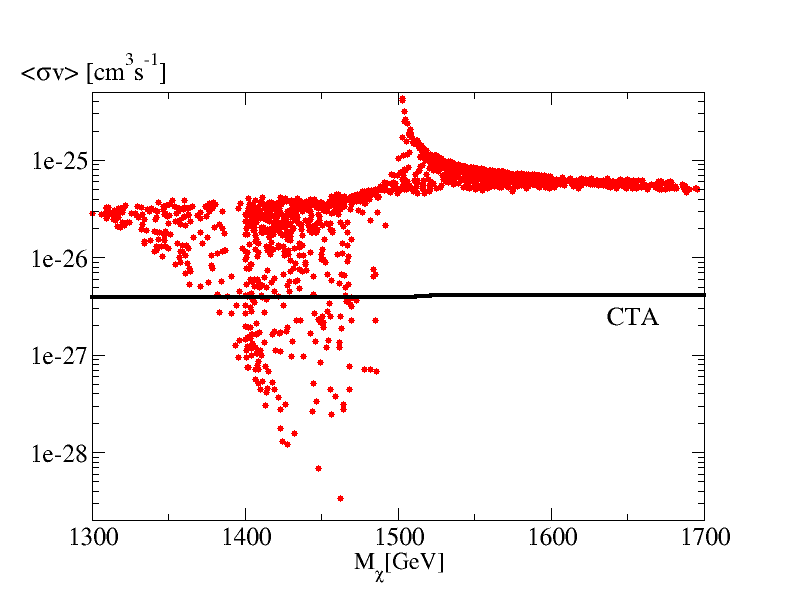}
	\caption{\textit{The expected self annihilation cross-section at zero velocity plotted as a function of mass. The black line represents an estimated cut from a future CTA experiment.\label{fig:indirect}}}
\end{figure}

Figure~\ref{fig:indirect} shows that the proposed limits exclude all points with $2m_\chi>\ma$ as in the zero velocity expansion of the self-annihilation cross section the momentum is mainly from the mass of the annihilating particles. Conversely for $2m_\chi<\ma$ a virtual mediator is produced.  There remain a section of points which evade all these bounds, due to the ability for the couplings to become extremely small near the s-channel resonance.

\section{Combined Experimental Constraints \label{combine}}
\subsection{Fixed $\ma$}
\begin{figure}[t]
	\center  
	\includegraphics[width=0.5\textwidth]{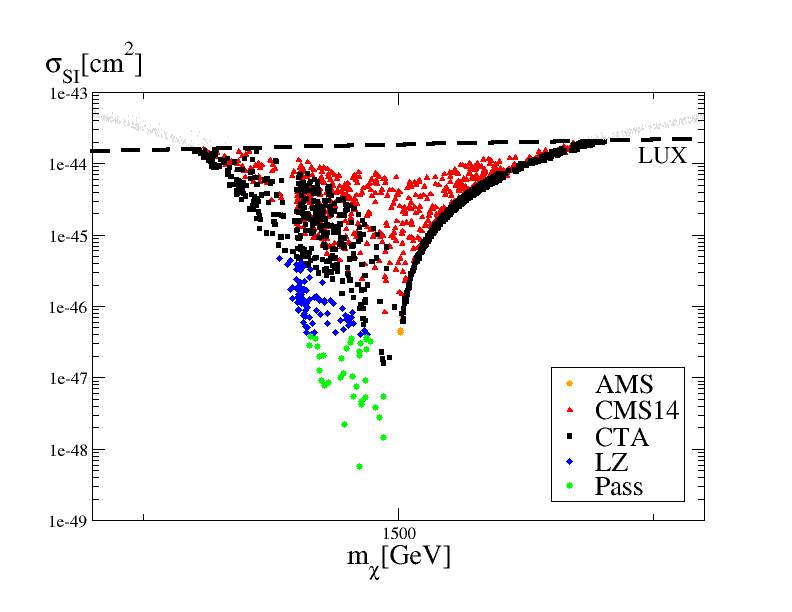}
\caption{\textit{Plots showing the effect of each search cut on the dark matter nucleon cross section parameter space allowed after the LUX cuts. Cuts are applied in the following order - Orange points pass LUX but are ruled out by AMS, red triangles pass AMS but are ruled out by dijets, black squares evade dijets but are ruled out by CTA, blue diamonds pass CTA but are ruled out by LZ, green points will not be ruled out by any of these experiments. \label{fig:combined}}}
\end{figure}
~
  \begin{figure}[t]
  \center
  \includegraphics[width=0.5\textwidth]{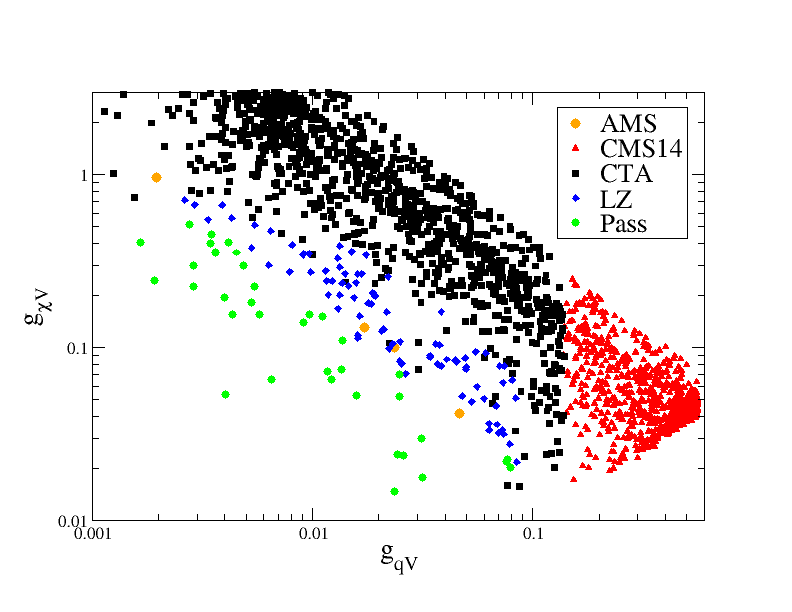}

	\caption{\textit{The same points at figure~\ref{fig:combined} plotted on the plane of the couplings.  \label{fig:gxgqfinal}}}
\end{figure}

From the last section it can be seen that a large volume of the possible parameter space can be excluded and that future experiments will explore quite varied kinematic and coupling regimes. Figures~\ref{fig:combined} and~\ref{fig:gxgqfinal} show the region that could be excluded by future experiments when we combine these constraints. In the mass-direct detection cross section plot (Fig.~\ref{fig:combined}) there is a region  localised on the right-hand side of the resonance curve of points that would not be seen by 14 TeV dijet searches but are excluded by CTA.  This is due to the fact that in this part of parameter space the decay of the mediator into $\chi\bar{\chi}$ is kinematically unfavourable in the Galactic centre. For these points in parameter space, the mediator will decay preferentially into standard model particles even for very small $g_{qV}$, including points with $g_{qV}$ too low to produce an observable dijet resonance signal.

The dijets would clearly be effective for setting an upper limit on the coupling to the standard model. The points which would not be excluded by these searches all lie in regions of parameter space where either both couplings are small or the coupling to dark matter is large which pushes down the quark coupling but finding a method to cut purely the dark matter coupling is extremely difficult. 

\subsection{Varied $\ma$}

In this section we allow $\ma$ to vary so as to see where the dark matter particles lie which cannot be excluded by any experiment.

\begin{figure}[H]
	\center  
	\includegraphics[width=0.5\textwidth]{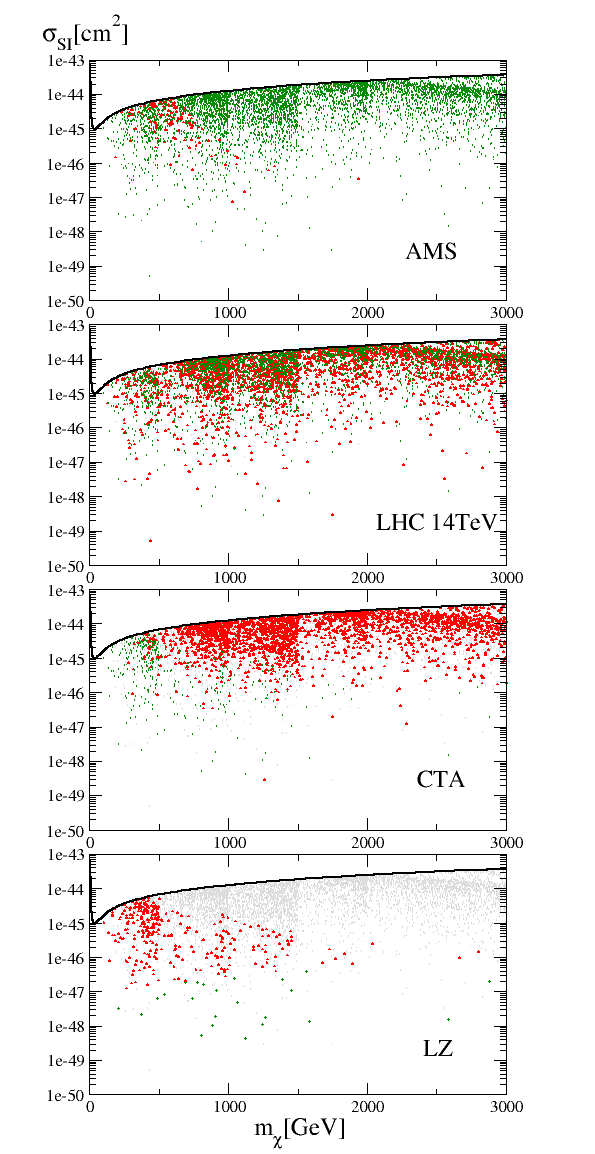}
\caption{\textit{Results for a range of different $\ma$ values, cuts are applied in order from top down, AMS, LHC, CTA then LZ.  In each plot, green points pass while red points are cut, grey points are ruled out by previous cuts.  Structure in the data simply comes from the search strategy having been split up into TeV mass ranges. \label{fig:VarMA}}}
\end{figure}

\begin{figure}[H]
  \center
  \includegraphics[width=0.5\textwidth]{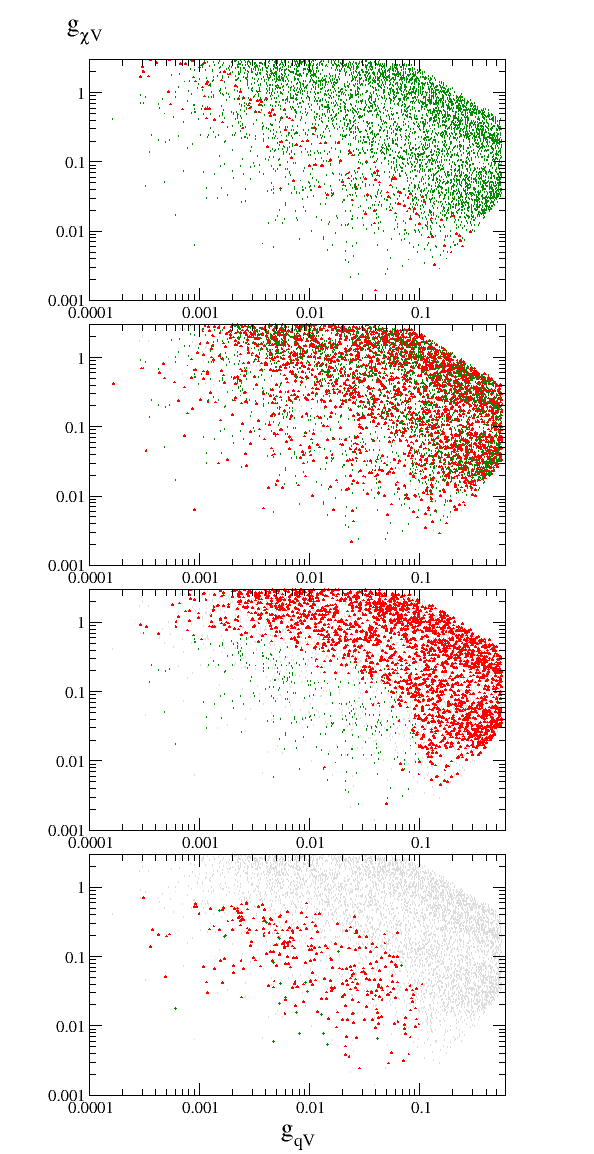}
  \caption{\textit{The same points at figure~\ref{fig:VarMA} plotted on the plane of the couplings. \label{fig:VarMA_couplings}}}
\end{figure}

 Figures~\ref{fig:VarMA} and~\ref{fig:VarMA_couplings} shows how the various experiments will restrict the parameter space for a range of $\ma$ around the resonance in the thermal relic density, the $m_\chi$ is allowed to vary from 0 to 3 TeV with $\ma$ being $~ 2m_\chi$. The plots show from the top down which points are ruled out and which are still acceptable.  The cuts are applied in order from the top down, AMS first, then dijets from a 14TeV LHC run then CTA then finally LZ.  In each plot, green points represent points which are still acceptable while red points are ruled out by that particular cut.  Grey points correspond to points that are ruled out by previous cuts.

Even for the lower mass dark matter the points which would not be excluded by LUX for this model still produce a monojet signature below the current experimental constraint.

\section{Conclusions}

There are many exciting new experimental programs which will shed light on the nature of dark matter over the next few years.  In this paper we have looked at the difficult situation where the dark matter mass is around half of the mediator mass.  In this situation, the resonant enhancement means that the correct relic abundance can be achieved with small couplings making direct detection difficult.

We have tried to use other techniques to rule out such models such as narrow dijet resonances and indirect detection and while these techniques do rule out many of the models which pass LUX, we still find that there are parameter values that give good dark matter candidates.  We also find that most of the models which can be ruled out using indirect detection are also ruled out by LZ.

There also exists a region of parameter space which would avoid direct detection where $m_\chi > \ma$ so that in the early universe $\chi\bar{\chi}$ could annihilate into two $A^\prime$ particles, but this requires a high $g_{\chi V}$ leading to points which have vanishingly low $g_{qV}$ so the constraints studied in this paper add no further cuts to that parameter space. 

From these results it is clear that a combined analysis of a model with comparisons to dedicated dark matter searches and effects of the given model can be used to greatly constrain the parameter space. While some of these searches cannot be used as a discovery claim as signal is seen (such as dijets) as many other models or affects could be responsible, they can be useful for cutting parts of the parameter space which would be acceptable if only the dedicated searches were taken into account. The use of the simplified model enables one to see clearly the dependence of different constraints on different model parameter values.  Hopefully this can inform possible constraints on more complete models.

\begin{acknowledgments}
This work was supported by the STFC.  We wish to thank John Paul Chou for advice on CMS dijet widths and Michael Gustafsson for comments on a draft version.
\end{acknowledgments}

\bibliography{Testbib}

\end{document}